\documentclass[12pt]{article}
 \usepackage[cp1251]{inputenc}

 \usepackage{amssymb}
 \usepackage{amsfonts}
 \usepackage{amsthm}
 \usepackage{amsmath}
 \usepackage{natbib}
 \usepackage{setspace}
 \usepackage{epsfig,color}
 \usepackage{multicol}
 \usepackage{hyperref}
 \usepackage{upgreek}
 \usepackage{longtable}
 \usepackage{caption}
 \usepackage[normalem]{ulem}
 \usepackage[english]{babel}

\title{A computer assisted proof for 100,000 years stability of the solar system}
\author{
  Angel Zhivkov$^1$\\
  \texttt{zhivkov@fmi.uni-sofia.bg} \and
  Ivaylo Tounchev$^1$\\
  \texttt{tounchev@fmi.uni-sofia.bg} }
\date{%
    $^1${Department of Mathematics and Informatics,
         Sofia University, 5 James Bourchier Blvd.,
         1164 Sofia, Bulgaria}\\[2ex]%
    \today
}

 \def\be  {\begin{equation}\begin{aligned}}
 \def\ee  {\end{aligned}\end{equation}}
 \def\ben {\begin{equation*}}
 \def\een {\end{equation*}}
 \def\ba  {\begin{align*}}
 \def\ea  {\end{align*}}

\begin{document}
 \label{firstpage}
  \maketitle

 \begin{abstract}
 We present an analytical proof assisted by computer calculations
 for the dynamical stability of the eight main planets and Pluto
 for the next 100,000 years.
 It means that the semi-major axes of the planets will not change
 significantly during this period. Also the eccentricities and
 inclinations of the orbits will remain sufficiently small.
 A standard linear four-step numerical method is used to integrate
 approximately the orbits of Mercury, Venus, Earth, Mars, Jupiter,
 Saturn, Uranus, Neptune and Pluto. Written in orbital elements, the
 dynamics of the nine planets manifests a system of 54 first-order
 ordinary differential equations. The step-size of the numerical
 method -- about six days, has been performed 6,290,000 times.
 We estimate the total accumulation of rounding-off errors,
 deviations related to possible uncertainty  in the astronomical
 data and the accuracy of the computer calculations.
 \end{abstract}


\section{Introduction}
 \label{Sec_Introduction}

    Important questions in celestial mechanics \citep{Hagihara} are:
 ``Will the present configuration of the solar system be preserved
   without radical changes for a long interval of time? What is the
   interval of time, at the end of which the
   configuration of the planetary orbits deviates from the present
   by a given small amount?"
  Next Hagihara claims that
  ``present mathematics hardly permit this question to be answered
   satisfactory for the actual solar system."

    On the other hand, after some simplifying assumptions in the
 secular perturbation theory of Laplace and Lagrange,
 the motion of the eight main planets -- from Mercury to Neptune,
 becomes integrable and the solar system would be stable.
 However, these simplifications are quite restrictive:
 one neglects the products of the planetary masses and assumes
 constant semi-major axes.

    In fact, it turns out that the exact dynamics of the planets
 is highly sophisticated. That is why almost all studies
 on the question use computers to carry out numerical integration
 either of the full equations of motion,
 or numerical integration of the averaged equations,
 see comments in \citep{MD}.
 Most of the computer simulations cover hundreds of millions or
 billions of years \citep{SW, Laskare}, but without any analytical
 evaluations about possible deviations of the obtained results
 from the real dynamics of the planets.

    Our algorithm for numerical integration is described in detail
 in section \ref{Code}.
 It is possible without substantial additional difficulties
 to improve considerably the accuracy of the numerical method.
 However this is not necessary because the biggest problems arise
 when one evaluates the deviations caused by small changes in the
 initial conditions, the masses of the planets or sun.

    For the calculations we use HP xw9400 workstation.
 The computer code performs 6,290,000 steps;
 each step-size is about six days, i.e. about 63 steps per year.
 Thus the integration covers 100,000 years.

  Our main result is the following

 \vspace{0.5ex}

 {\bf Theorem.} {\it The configuration of the osculating ellipses on
 which the planets move around the Sun will remain stable at least
 100,000 years in the sense that the  semi-major axis of each planet
 varies within or less than $1 \% $. The maximal values of the
 eccentricities and inclinations of the orbits remain bounded.}

 \vspace{0.5ex}

 In the conditions of the Theorem, we also suppose possible
 uncertainties in the data from astronomical observations:
 up to $\pm 10^{-4}$ absolute error in the orbital elements
 and the masses of the  planets for the J2000 epoch.
 By exception only for the masses of Mercury, Mars and Pluto
 we assume up to $\pm 10\%$ possible relative deviations.


   The paper has the following structure.

   In Section \ref{Equations} we define three sets of
 generalized co-ordinates and the equations of motion,
 corresponding to the third set.

   In section \ref{Code} we define additional helpful
 variables to enable the work of the computer code.
 Certain details for this code are discussed.

 In section \ref{Results} we discuss the results
  of the numerical integration.
  The maximal deviations of the semi-major axes, eccentricities
  and inclinations of the orbits are listed.

   In section \ref{Proof} we prove that for any orbital element the
 error of the numerical integration does not exceed $10^{-9}$ per
 step.
 Thus the final result would not be differ significantly
 from the real dynamics of any planet.

   Next we change slightly one by one the 54 initial conditions
 and the masses of the Sun and the planets.
 The obtained numerical results show that the solar system dynamics
 is stable under sufficiently small changes of initial conditions,
 masses and sufficiently small additional perturbations.

 In section \ref{Conclusions} we make our conclusions.

 \section{Equations of motion}
 \label{Equations}

  In a heliocentric coordinate system, any planetary orbit
  can be described by six time-depending functions.
  We shall need three different such six-tuples:

  \vspace{0.1ex}
  $(x,y,z,\dot{x},\dot{y},\dot{z})$
  -- the usual rectangular coordinates and velocities in
  ${\mathbb R}^3$, dot denotes a differentiation with respect
  to the time $t$,

  \vspace{0.1ex}
  $(a,e,i,l,g,\theta)$
  -- six orbital elements,
  $a$ is the semi-major axis,
  $e$ is the eccentricity,
  $i$ is the inclination of the orbit,
  $l$ is the mean anomaly,
  $g$ is the argument of perihelion and
  $\theta$ is the longitude of ascending node,

  \vspace{0.1ex}
  $(a,\epsilon,h,k,p,q)$
  -- an alternative system of six orbital elements,
  $\epsilon$ is the modified mean longitude at epoch \citep{BC},
  $h$ and $k$ are eccentric elements,
  while $p$ and $q$ are oblate elements.
\vspace{0.1ex}
  The first and the second coordinate systems
  have been related by \citep{Poincare}

  \be
   \label{xyz}
   x\, & =(\cos g \cos\theta - \sin g \sin\theta \cos i)\, X
         -(\sin g \cos\theta + \cos g \sin\theta \cos i)\, Y \,,
                                                                 \\[4pt]
   y\, & =(\cos g \sin\theta + \sin g \cos\theta \cos i)\, X
         -(\sin g \sin\theta - \cos g \cos\theta \cos i)\, Y \,,\
                                                                 \\[4pt]
   z\, & = \sin i\, \sin{g} \ X \,+\, \sin i\, \sin{g}  \  Y \,,
 \ee
 where \citep{AW}
 \be
  \label{XY}
    X &= a\,\biggl[ \cos l - e -
         \sum\limits_{s=1}^\infty
          \frac{e^s}{s!} \frac{d^{s-1} \sin^{s+1}l}
                              {dl^{s-1}}\biggr]
                                             \ , \\[0.5ex]
    Y &= a\,\sqrt{1-e^2}
         \biggl[ \sin l
              +\sum\limits_{s=1}\frac{e^s}{s!}
               \frac{d^{s-1} (\sin^s l \cos l)}{dl^{s-1}}
         \biggr]  \ ,
 \ee
 and the mean anomaly $l$ and the mean motion $n$
 are connected by the Kepler's third law:
 \be
   \label{lnmu}
     l & \,= n \,(t-t_0)                        \,,\\[3pt]
     n &  := \sqrt{1+\mu}\, a^{-3/2}            \,,\\[3pt]
   \mu &  := \frac{\textrm{mass of the planet}}
                  {\textrm{mass of the Sun}}    \,,
 \ee
 $t_0$ is a time of perihelion passage.

    Analogous to (\ref{xyz}) equations hold for
 $\dot{x}$, $\dot{y}$ and $\dot{z}$, just replace
 $X$ by $n\,\partial_l X$ and $Y$ by $n\,\partial_l Y$.
 By $\partial_w$ we shall denote the
 partial derivative with respect to the variable $w$.

 On the other hand,
 \vspace{-1ex}
 \be
  \label{hkpq}
   \lambda & := l+g+\theta         \,,\quad
     \epsilon(t) := \lambda(t)-\int_{0}^{t}n(s) \,ds   \,, \\[2pt]
         h & := e \sin(g+\theta)   \,,\quad
         k   := e \cos(g+\theta)   \,, \\[5pt]
         p & := 2 \sin\tfrac{i}{2} \sin\theta  \,,\quad
         q   := 2 \sin\tfrac{i}{2} \cos\theta
 \ee
 relates the above two systems of orbital elements.
 The mean longitude $\lambda$ is a fast variable while the modified
 mean longitude at epoch $\epsilon$ is a slow variable.

   We shall consider the Sun and planets
 as ten point masses, indexed by $j=0,1,2,\ldots,9$,
 located at their position vectors ${\bf r}_j$
 and moving with 3-velocity $\dot{\bf r}_j$,
 \ben
       {\bf r}_j := \big( x_j,y_j,z_j \big)^t         \,, \qquad
   \dot{\bf r}_j := \big(\dot{x}_j,\dot{y}_j,\dot{z}_j\big)^t \ .
 \een
 Choose the unit of mass to be the mass of the Sun
 and the unit of distance to be the mean distance between the
 Earth and the Sun, i.e. one astronomical unit (AU).
 Then the relative mass $\mu_j$ of the $j$th planet satisfies
 (\ref{lnmu}) and time $t=2\pi$ corresponds to one Julian year,
 i.e. $365.25$ days. Time $t=0$ corresponds to the year A.D. $2000$.

   According to the Newton's theory of gravitation,
 the 3-accelerations of the planets
 \ben
     \ddot{\bf r}_j  =  -(1+\mu_j)\frac{{\bf r}_j}{r_j^3}
                      \,+\, \partial_{{\bf r}_j} R_j
                   \, ,\qquad \ j=1,\ldots, 9 \ .
 \een
 Here $r_j^2=x_j^2+y_j^2+z_j^2$ is the squared distance
 between the Sun and $j$th planet.
 The disturbing function
 \be
  \label{R}
   R_j & :=\sum\limits_{s \not= j}
            \mu_s \biggl[\frac{1}
                            {({\bf r}_j {\bf r}_j
                             +{\bf r}_s {\bf r}_s
                            -2{\bf r}_j {\bf r}_s)^{1/2}}
                       -\frac{{\bf r}_j {\bf r}_s}
                             {r_s^3}
                  \biggr]                              \ ,
 \ee
 where ${\bf r}_b {\bf r}_c = x_b x_c+y_b y_c + z_b z_c\,$
 is the Euclidean dot product.
 Also, the gradient
 \ben
       \partial_{{\bf r}_j} R_j
     = \big( \partial_{x_j} R_j\,,\,
             \partial_{y_j} R_j\,,\,
             \partial_{z_j} R_j
       \big)^t                                   \ .
 \een

   Now we apply the classical Lagrange's brackets transform
 \citep{BC} from  coordinates $(x,y,z,\dot{x},\dot{y},\dot{z})$
 to $(a,\epsilon,h,k,p,q)$. This leads to the system of
 $9\!\cdot\! 6 = 54$ first order ODE's
 \be
  \label{EQ}
   \dot{a} & =  +\frac{2}{na}\,\partial_\epsilon R  \,, \\[4pt]
   \dot{\epsilon} & =  - \frac{2}{na}\,\partial_a R
                    + \frac{ h\,\partial_h R + k\,\partial_k R}
                           {na^2 (1+E^{-1})}
                    + \frac{ p\,\partial_p R +q\,\partial_q R}
                           {2na^2 E}        , \\[4pt]
   \dot{h} & = \,+\,\frac{E}{na^2}\,\partial_k R
                - \frac{ h\,\partial_\epsilon R}
                       {na^2(1+E^{-1})}
                + k\,\frac{ p\,\partial_p R + q\,\partial_q R}
                          {2na^2 E} \,, \\[4pt]
   \dot{k} & = \,-\,\frac{E}{na^2}\,\partial_h R
                - \frac{ k\,\partial_\epsilon R}
                       {na^2(1+E^{-1})}
                - h\,\,\frac{ p\,\partial_p R + q\,\partial_q R}
                            {2na^2 E} \,, \\[4pt]
    \dot{p} & = \, +\,\frac{E^{-1}}{na^2}\,\partial_q R
                  -p\,\frac{     \partial_\epsilon R
                            + k\,\partial_h R - h\,\partial_k R}
                           {2na^2 E}  \,,    \\[4pt]
    \dot{q} & = \, -\,\frac{E^{-1}}{na^2}\,\partial_p R
                  -q\,\frac{     \partial_\epsilon R
                            + k\,\partial_h R - h\,\partial_k R}
                           {2na^2 E}  \,,
 \ee
 where $E:=\sqrt{1-h^2-k^2}$.
 For simplicity, all the running from 1 to 9 indexes $j$
 have been omitted and one should understand
 $a_j$ instead of $a$, $\epsilon_j$ instead of $\epsilon$, ... ,
 $R_j$ instead of $R$.

 As usual, in the expression (\ref{R}) for the disturbing
 function $R_j$ we use the mean longitudes $\lambda_j$ or
 $\lambda_s$  instead of the modified longitudes at epoch
 $\epsilon_j$ or $\epsilon_s$.
 In the equations of motion (\ref{EQ}), under differentiating
 with respect to $\epsilon$ we understand differentiating
 with respect to $\lambda$ viz.,
 \ben
      \partial_{\epsilon_j}=\partial_{\lambda_j} \,, \quad
      \partial_{\epsilon_s}=\partial_{\lambda_s} \,.
 \een

    Note a technical difference in (\ref{EQ}) with respect to
 the equations of motion used in \citep{BC} or \citep{MD}.
 Namely, we prefer the multiplier $2\sin\frac{i}{2}$
 when define the oblate elements $p$ and $q$ in (\ref{hkpq}),
 in contrast to the corresponding multipliers
 $\tan i$ or $\sin i$.

 \section{Numerical code}
 \label{Code}

   In order to solve numerically the system of 54 ODE's (\ref{EQ})
 we apply the simplest four-step method \citep{Butcher}.
 For any variable $\sigma=a_1,\lambda_1,\ldots,p_9$ or $q_9$
 the iteration formula reads
 \be
  \label{iteration}
     \sigma(\tau s) & \approx\,  \sigma_s                 \,,  \\
     \sigma_{s+1}   & := \sigma_s + \frac{\tau}{24}\,
                          \big( 55\dot{\sigma}_s
                               -59\dot{\sigma}_{s-1}
                               +37\dot{\sigma}_{s-2}
                               -9 \dot{\sigma}_{s-3}
                          \big)
 \ee
 for $s$ from $0$ to $6,290,000$ and step-size
 \ben
     \tau = 0.1   \,.
 \een
 The error $|\sigma(\tau s) - \sigma_s|= O(\tau^5)$ will be
 specified in section 5.

    For initial conditions in (\ref{EQ}) we cite \citep{St, NASA}.
 Before starting the iterations with $s=0$,
 it is necessary to carry out $300$ single backward Euler steps
 with step-sizes $\tau=10^{-3}$ to find $\dot{\sigma}_{-1}$,
 $\dot{\sigma}_{-2}$ and $\dot{\sigma}_{-3}$.
 Approximately, each step-time equals to six days, $63$ steps
 cover one year and $6,290,000$ steps cover more than $100,000$ years.

   The numerical code is organized as follows.

  First we define the intermediate eccentric variables
 \be
  \label{HK}
    H &:= e \sin l =    -h \cos\lambda + k \sin\lambda \,,\\[3pt]
    K &:= e \cos l = \ \ h \sin\lambda + k \cos\lambda \,.
 \ee
 and calculate the expansions (\ref{XY}) of $X$ and $Y$
 up to order $\leq 10$ in the eccentricity $e$.
 It turns out that the new variables
 \begin{align*}
    U &:= (X \cos l + Y \sin l) \cos\lambda
         +(X \sin l - Y \cos l) \sin\lambda         \\[4pt]
      &=  a \big( 1-K-H^2 + \cdots\big) \cos\lambda
         + aH \big( 2 + \tfrac{1}{2}K + \cdots \big)\sin\lambda
             +Err_{11}(U) \,
 \end{align*}
 and
 \begin{align*}
    V &:= (X \cos l + Y \sin l) \sin\lambda
         -(X \sin l - Y \cos l) \cos\lambda         \\[4pt]
      &=  a \big( 1-K-H^2 + \cdots\big) \sin\lambda
          - aH\big( 2 +\tfrac{1}{2}K + \cdots \big)\cos\lambda
             +Err_{11}(V)
 \end{align*}
 do not depend explicitly on the mean anomaly $l$.
 Inside the $Err_{11}(U)$ and $Err_{11}(V)$ are included all terms
 of degree $\ge 11$ in $H$ and $K$.

   Neglecting the eleventh and  higher order eccentricities
 in the code, the approximate coordinates $x$ and $y$ become
 polynomials of degree twelve:
 \begin{align*}
     x &= \big(1-\tfrac{1}{2}p^2 \big) U + \tfrac{1}{2}pq V  \,,\\[4pt]
     y &= \big(1-\tfrac{1}{2}q^2 \big) V + \tfrac{1}{2}pq U  \,,\\[4pt]
     z &= \big[1-\tfrac{1}{4}(p^2+q^2)\big]^{1/2} (-pU+ qV)  \ .
 \end{align*}
 Remark that these formulae look much convenient compared
 with (\ref{xyz}).
 In view of (\ref{HK}) one can easily calculate all the derivatives
 of $x$, $y$ and $z$ with respect to $\lambda$, $h$ and $k$.
 Also, there is no problem to take derivatives with respect to
 $a$, $p$ or $q$.

    Let us now briefly explain the structure of our numerical code.
    Each $\tau$-step requires as initial conditions five
 $(9 \times 6)$-matrices $S$, $dS$, $dS\_1$, $dS\_2$ and  $dS\_3$.

    The matrix $S$ fixes the numerical values of
 $a_j,\epsilon_j,h_j,k_j,p_j$ and $q_j$ at its $j$th row.
 In terms of (\ref{iteration}), each entry of $S$ corresponds
 to $\sigma_s$.

    On the other hand, the matrices $dS$, $dS\_1$, $dS\_2$
 and $dS\_3$ remember the values of the derivatives
 \ben
     \dot{a}_j,\,\dot{\epsilon}_j,\,\dot{h}_j,
     \,\dot{k}_j,\,\dot{p}_j,\,\dot{q}_j,
     \qquad 1 \leq j \leq 9
 \een
 at the $s$th, $(s-1)$th, $(s-2)$th and $(s-3)$th steps
 correspondingly.
 In terms of (\ref{iteration}) these last four
 matrices define $\dot{\sigma}_{s}$, $\dot{\sigma}_{s-1}$,
 $\dot{\sigma}_{s-2}$ and $\dot{\sigma}_{s-3}$.

    We define three procedures to perform the $s$th step.
    The procedure ``make\_xyz'' computes $x,y,z$ and their
 derivatives with respect to $a,\epsilon, h,k,p,q$
 with the help of the above defined additional variables
 $K,H,U$ and $V$.

    The procedure ``make\_2'' calculates all dot products
 \begin{align*}
    & {\bf r}_j {\bf r}_{s}\,,
    (\partial_{a_j}{\bf r}_j) {\bf r}_{s}\,,
    (\partial_{\lambda_j}{\bf r}_j) {\bf r}_{s}\,,
    (\partial_{h_j}{\bf r}_j) {\bf r}_{s}\,,
    (\partial_{k_j}{\bf r}_j) {\bf r}_{s}\,,
    (\partial_{p_j}{\bf r}_j) {\bf r}_{s}\,,
    (\partial_{p_j}{\bf r}_j) {\bf r}_{s}\,,
 \end{align*}
 for $j$ and $s$ ranging from 1 to 9.
  Next ``make\_2'' computes all mutual distances
 $\Delta_{j,s}$ between the planets:
 \begin{align*}
    \Delta_{j,s}^2 &=    {\bf r}_j {\bf r}_{j}
                     +   {\bf r}_s {\bf r}_{s}
                     -2\,{\bf r}_s {\bf r}_{s} \,,
 \end{align*}
 and finally computes the partial derivatives
 \begin{align*}
  \partial_{a_j} R_j \,,\,
  \partial_{\lambda_j} R_j\,,\,
  \partial_{h_j} R_j \,,\,
  \partial_{k_j} R_j \,,\,
  \partial_{p_j} R_j \,,\,
  \partial_{q_j} R_j
 \end{align*}
 of the disturbing functions (\ref{R}).

   Given $54$ values of these last derivatives,
 the third procedure ``make\_EQ'' finds a
 $(9\times 6)$--matrix $dS$ of the new values of the velocities
 $\dot{a}_j,\dot{\epsilon}_j,\dot{h}_j ,
   \dot{k}_j,\dot{p}_j,\dot{q}_j$
 according to the equations of motion (\ref{EQ}).

    Finally, every single step executes:

 (i) the procedure ``make\_xyz'',

 (ii) the procedure ``make\_2'',

 (iii) the procedure ``makeEQ'',

 (iv) the numerical scheme
 \begin{align*}
     S &:= S + \frac{0.1}{24} \big( 55\, dS -59\, dS\_1
             +37\, dS\_2-9\, dS\_3  \big) \ ,
 \end{align*}
 which defines the new values of
 $a_j, \epsilon_j, h_j, k_j, p_j$ and $q_j$,

 (v) compute the new mean longitudes $\lambda$:
 \begin{align*}
    \lambda_{s+1} &:= \lambda_s + \epsilon_{s+1}
                                - \epsilon_s
     + \frac{0.1}{24}
             \Big(  \frac{55}{2}\dot{n}_s
                  - \frac{59}{3}\dot{n}_{s-1}
                  + \frac{37}{4}\dot{n}_{s-2}
                  - \frac{9}{5} \dot{n}_{s-3}
             \Big),
 \end{align*}
 see the second equation in (\ref{hkpq}),

 (vi) $dS$ becomes $dS\_1$\,, $dS\_1$ becomes $dS\_2$\,,
     $dS\_2$ becomes $dS\_3$\,.

  After $6,290,000$ times repeating these (i)--(vi) points
  we obtain the final matrix $S$ and so the approximated
  position of each planet after $100,000$ years.

 Remark: The implemented code for numerical
 integration in the present paper can be made
 available upon request to the  authors.

 \section{Results and discussion}
 \label{Results}

 The numerical integration of the equations of motions
 (\ref{EQ}) for the next $10^{5}$ years shows
 more or less expected results.

   The semi-major axes $a_j$ oscillate within the
 periodic perturbations, see the next table.

 \begin{center}
 \captionof{table}{Minimum and maximum values of the
 semi-major axis (in AU).
 Displacement (in degrees) of the modified mean longitude
 at epoch $\Delta \epsilon$ for $10^5$ years.
 Mean motions $n$ for the next $10^5$ years.}
 \begin{tabular}{||c c c c c||}
 \hline
 \hline
         &$a_{min}$ & $a_{max}$ & $\Delta\epsilon$ & $n$ \\[0.5ex]
 \hline
 Mercury & 0.3870 & 0.3871 & $  -187^{\circ}$ & 4.1524 \\
 \hline
 Venus   & 0.7233 & 0.7234 & $  -288^{\circ}$ & 1.6258 \\
 \hline
 Earth   & 1      & 1.0001 & $\ \ 50^{\circ}$ & 1.0001 \\
 \hline
 Mars    & 1.5232 & 1.5238 & $  -237^{\circ}$ & 0.5318 \\
 \hline
 Jupiter & 5.1997  & 5.2038  & $\  -14^{\circ}$ & 0.0844 \\
 \hline
 Saturn  &  9.4670 & 9.5426  & $  3122^{\circ}$ & 0.0342 \\
 \hline
 Uranus  & 19.034 & 19.280 & $  1177^{\circ}$ & 0.0120 \\
 \hline
 Neptune & 29.766 & 30.247 & $\  599^{\circ}$ & 0.0061 \\
 \hline
 Pluto   & 38.614 & 40.116 & $\  408^{\circ}$ & 0.0041 \\
 \hline\hline
\end{tabular}
\end{center}

 The small oscillations $a_{1,2,3,4,5}$ of the inner
 planets and Jupiter are consequence of the choice
 of heliocentric reference frame.
 The relatively large oscillations of $a_{6,7,8,9}$
 are due to the indirect parts
 \ben
    -\sum\limits_{s < j}
          \mu_s \frac{{\bf r}_j {\bf r}_s}
                      {r_s^3}
 \een
 of the perturbations $R_j$ combined with the following
 near resonant mean motions:
 \begin{align*}
   & n_5:n_6 \approx 5:2  \,,\quad
     n_6:n_7 \approx 20:7 \,,\quad
     n_7:n_8 \approx 2:1  \,,\quad
     n_8:n_9 \approx 3:2  \,.
 \end{align*}

 The mean motions $n_j$ for the next $10^5$ years remain
 almost the same as to ``instant mean motions $n_j$
 at the epoch $J2000$'', see $4$th column of table 1.

 The eccentric and oblate variables manifest some unanticipated
 results, see the next two tables.

 \newpage
 \begin{center}
 \captionof{table}
 {Minimum and maximum values of the eccentricities.
 Initial and final values of
 the longitude of perihelion.}
 \begin{tabular}{||c c c c c||}
 \hline\hline
       & $e_{min}$ & $e_{max}$ & $\tilde{\omega}_{2000}$
       & $\tilde{\omega}_{6 \cdot 10^{5}}$                \\ [0.5ex]
 \hline
 Mercury & $0.195$ & $0.208$ & $\ 77^{\circ}$ &  $222^{\circ}$ \\
 \hline
 Venus   & $0.002$ & $0.022$ & $ 132^{\circ}$ &\ $98^{\circ}$  \\
 \hline
 Earth   & $0.002$ & $0.017$ & $ 103^{\circ}$ &  $476^{\circ}$ \\
 \hline
 Mars    & $0.086$ & $0.121$ & $ -24^{\circ}$ &  $471^{\circ}$ \\
 \hline
 Jupiter & $0.025$ & $0.060$ & $\ 15^{\circ}$ &  $160^{\circ}$ \\
 \hline
 Saturn  & $0.010$ & $0.088$ & $\ 92^{\circ}$ &  $863^{\circ}$ \\
 \hline
 Uranus  & $0.028$ & $0.055$ & $ 171^{\circ}$ &  $280^{\circ}$ \\
 \hline
 Neptune & $0.001$ & $0.018$ & $\ 45^{\circ}$ &\ $ 83^{\circ}$ \\
 \hline
 Pluto   & $0.237$ & $0.266$ & $ 224^{\circ}$ &  $226^{\circ}$ \\
 \hline\hline
\end{tabular}
\end{center}

\begin{center}
\captionof{table}{Minimum and maximum values of the inclination.
    Initial and final values of the longitudes of ascending node.}
\begin{tabular}{||c c c c c||}
 \hline\hline
       & $i_{min}$  & $i_{max}$  & $\theta_{2000}$ & $\theta_{6 \cdot 10^{5}}$  \\ [0.5ex]
 \hline
 Mercury & $5.8^{\circ}$ & $7^{\circ}$ & $48^{\circ}$ & $-143^{\circ}$ \\
 \hline
 Venus & $0.05^{\circ}$ & $3.4^{\circ}$ & $77^{\circ}$ & $-882^{\circ}$  \\
 \hline
 Earth & $0^{\circ}$ & $3.3^{\circ}$ & $0^{\circ}$ & $-244^{\circ}$  \\
 \hline
 Mars & $0.4^{\circ}$ & $4.5^{\circ}$ & $50^{\circ}$ & $-561^{\circ}$  \\
 \hline
 Jupiter & $1.2^{\circ}$ & $2^{\circ}$ & $101^{\circ}$ & $109^{\circ}$  \\
 \hline
 Saturn & $0.7^{\circ}$ & $2.6^{\circ}$ & $114^{\circ}$ & $97^{\circ}$  \\
 \hline
 Uranus & $0.6^{\circ}$ & $1.4^{\circ}$ & $74^{\circ}$ & $151^{\circ}$  \\
 \hline
 Neptune & $0.05^{\circ}$ & $3.4^{\circ}$ & $132^{\circ}$ & $125^{\circ}$  \\
 \hline
 Pluto & $16.6^{\circ}$ & $17.2^{\circ}$ & $110^{\circ}$ & $95^{\circ}$  \\ [1ex]
 \hline\hline
\end{tabular}
\end{center}

 In addition to the above arguments for the indirect parts and
 near-resonances there are two main reasons for the discrepancy
 with the Laplace--Lagrange secular theory,
 as well as with the Brouwer \& van Woerkom
 secular theory \citep{BW}.
 In contrast of the numerical integration these theories does
 not take into account the fourth order (in the eccentricities
 and inclinations) secular terms.

 Another effect arises when some eccentricity or inclination
 is very small. Then the corresponding longitude of perihelion
 or node becomes chaotic behavior, with possible $\pm 360^{\rm o}$
 errors in their numerical integration.

 For instance, it turns out that the longitude of perihelion
 of Venus for the next $100,000$ years moves clockwise
 from $132^{\rm o}$ to $90^{\rm o}$.
 Another examples are the counterclockwise movements of
 Jupiter and Uranus nodes.

\section{Proof of the Theorem}
 \label{Proof}

 Let ${\bf u}^*$ be the explicitly obtained after numerical
 integration of the equations of motion (\ref{EQ})
 approximate solution.
 Suppose also that ${\bf v}={\bf v}(t)$ be some exact
 solution, which initial conditions  ${\bf v}(0)$
 are close to ${\bf u}^*(0)$.

   Then the proof of the theorem requires to show that the
 deviation
 \be
  \label{dev}
    {\boldsymbol\delta}(t) := {\bf v}(t)-{\bf u}^\ast(t)
 \ee
 will be sufficiently small for the next $100,000$ years.

    So, we shall represent and estimate $ {\boldsymbol\delta}$
 as a superposition of the error of the numerical integration
 method and the linear and non-linear variations due to
 the difference ${\bf v}(0)\not= {\bf u}^*(0)$.

 \subsection{The maximal error of a single step}
 \label{error}

    The calculations are made with accuracy of $10^{-10}$ for
 each step of the algorithm, which gives an accumulation
 less than $10^{-4} $ for the entire period.

    Also, the estimation of the errors
 \ben
    \left |Err_{11}(U)\right | + \left |Err_{11}(V)\right |
   \leq \frac{e^{11}}{1-e} \exp\frac{11}{2}
 \een
 follows from an estimation of the derivatives in (\ref{XY})
 for $j \ge 11$.
 In the calculations, we have ignored these errors which
 gives an error less than $10^{-10}$ per step.
 With accumulation this would yield deviations less than
 $6 \cdot 10^{-4}$ in the final values of the action
 variables $a$, $e$ and $i$.

    Next, let $\sigma$ be an element of the $j$th planetary orbit,
 that is $\sigma \in \{a_j,\epsilon_j,h_j,k_j,p_j,q_j \}$.
 During the procedure of numerical integration of $\sigma$
 we have neglected the fifth time--derivative $\sigma^{(5)}(t)$
 and now the error coming from the numerical method will be estimated.
 It will be done for every $\tau$-step and afterwards accumulated
 for all $6,290,000$ $\tau$-steps.

    Thus one should apply the operator
 \be
   \label{oper}
    &\frac{d}{dt} \ = \ \negthickspace\sum_{m=1}^{9}
                        \big(\dot{a}_m \partial_{a_m}\!
                +\dot{\epsilon}_m \partial_{\epsilon_m}\!
                +\dot{h}_m\partial_{h_m}\!
                +\dot{k}_m \partial_{k_m}
                +\dot{p}_m \partial_{p_m}
                +\dot{q}_m \partial_{q_m}\big)
 \ee
 four times over the  $\dot{\sigma}$ right side in (\ref{EQ}).
 Since $\dot a, \dot h, \dot k, \dot p$ and $\dot q$ are of order
 $10^{-3}$ (or even smaller) and $\dot{\epsilon}=\dot{\lambda}+n$,
 we can consider that
 \ben
    \frac{d}{dt}\approx \sum_{m=1}^{9}n_m \partial_{\lambda_m}
 \een
 and consequently
 \be
   \label{appr}
    \frac{d^4}{dt^4}\approx
    \sum_{m=1}^{9}\sum_{b=1}^{9}
    \sum_{c=1}^{9}\sum_{d=1}^{9}
       n_m n_b n_c n_d \,
       \partial_{\lambda_m}\partial_{\lambda_b}
       \partial_{\lambda_c}\partial_{\lambda_d}\,.
 \ee
 It is easily to see that the approximated $\frac{d^4}{dt^4}$ differs
 from the exact fourth degree of $\frac{d}{dt}$ from (\ref{oper})
 with no more than $10\%$.

    The right side of (\ref{appr}) is a sum of $9^4$ differential
 operators but only $129$ of them do not nullify the right side of
 the equation for $\dot{\sigma}$. These are those for which among the
 indexes $j, m, b, c, d$ no more than two are different, see the
 equation (\ref{R}) for the disturbing function.

 The mean motions $n_{5,6,7,8}$ of the big planets are relatively
 small and so the differentiations with respect to
 $\lambda_{5,6,7,8}$ compensate the relatively large masses
 $\mu_{5,6,7,8}$,
 \begin{align*}
     \mu_{1,2,\ldots,9} \approx
     & \bigg[\frac{1}{6023600},
             \frac{1}{408523},
             \frac{1}{328900},
             \frac{1}{3098708}, \\[2pt]
     & \ \   \frac{1}{1047.35},
             \frac{1}{3497.89},
             \frac{1}{22903},
             \frac{1}{19412},
             \frac{1}{1.35 \cdot\! 10^8}
       \bigg]  \,.
 \end{align*}
 Reversely, the large values of $n_{1,2,3,4}$ are compensated by the
 small values of $\mu_{1,2,3,4}$. The perturbations caused by Pluto
 are entirely negligible.

    Straightforward calculations show that for all $\sigma$
 \begin{align}
  \label{sigma5}
   \big| \sigma^{(5)}(t) \big| <  \frac{1}{200}.
 \end{align}
 Such an estimate takes place only for $\sigma=\epsilon_1$.
 For the remaining $53$ orbital elements the evaluation is
 $|\sigma^{(5)}|<\frac{1}{1000}$.

     Taking into account the coefficient $\frac{0.1^5}{5!}$ in front
 of the remainder in Taylor series expansion of $\sigma$ we conclude
 that the exponentially accumulated error \citep{Butcher} of the
 numerical integration  for $10^5$ years does not exceed
 \begin{align*}
  \label{err}
     \biggl[ 1+\frac{0.1^5}{5!} \frac{1}{200}
     \biggr]^{6290000} \!\! - 1
    \ < \ \frac{1}{380}              \,.
 \end{align*}

   In practice, the real estimate is much better since the signs of
 $\sigma^{(5)}(t)$ do alternate for different $t$ and largely
 compensate each other.

\subsection{Variations due to different initial conditions}
 \label{Ini_err}

 Let us extend (\ref{EQ}) with nine new differential equations
 $\dot{\mu_j}=0$ and rewrite this new system of $63$ ODE's as
 \ben
  \label{ext}
    {\bf \dot{v}} = {\bf f}({\bf v}) \,,\quad
    {\bf v}:=(a_1, \epsilon_1,..,q_1,\mu_1,\ldots,
     a_9,\epsilon_9,..,q_9,\mu_9) \,.
 \een

   Introduce next a metric
 \be
  \label{vnorm}
    &||{\bf v}|| := \max \Big( a_1,a_2,a_3,a_4,
                              \frac{a_5}{5},
                              \frac{a_6}{10},
                              \frac{a_7}{20},
                              \frac{a_8}{30},
                              \frac{a_9}{40},
  \\[4pt] & \hspace{2.6cm}
                       \max\limits_{1\leq j \leq 9}
                       \big( 0.3 |\epsilon_j|,|h_j|,|k_j|,|p_j|,|q_j| \big),
  \\[0pt] & \hspace{2.6cm}
                      10^5 \mu_1, \mu_2,\mu_3,10^5 \mu_4,
                      \mu_5,\mu_6,\mu_7,\mu_8,10^5 \mu_9
    \Big),
 \ee
 where all $\epsilon_j \in (-\pi,\pi]$.
 Roughly speaking, the solutions of (\ref{EQ}) do
 not exceed $2$ by norm.

 We shall suppose that the varied initial conditions
 \ben
    ||{\boldsymbol\delta}(0)|| = ||{\bf v}(0)-{\bf u}^\ast(0)||
    < 10^{-4} .
 \een
 The explicitly obtained numerical solution ${\bf u}^*$ has initial
 conditions ${\bf u}^*(0)$ set in \citep{St, NASA} and satisfies a
 system of $63$ ODE's
 \ben
      {\bf \dot{u}}^\ast = {\bf f}^*({\bf u}^\ast) \ ,
 \een
 where the vector function ${\bf f}^*\approx {\bf f}$.

   To be precise, as consequence of (\ref{sigma5})
 \be
    \label{fsf}
      ||{\bf f}^*({\bf v})-{\bf f}({\bf v})||
      < \frac{6}{5}\frac{10^{-5}}{5!}\frac{1}{200}
      =5\cdot 10^{-10}
 \ee
 provided
 \be
    \label{sigma}
       ||\boldsymbol\delta(t)||
     = ||{\bf v}(t)-{\bf u}^*(t)|| < 10^{-3} \quad
       \textrm{for all} \ \ t \leq 2\pi \!\cdot\! 10^5 .
 \ee
 Indeed, the velocities $\dot{a}_1, \ldots, \dot{q}_9$ in
 (\ref{EQ}) are small as well as their first partial
 derivatives and so the Taylor's expansion of $\bf f$
 says that a multiplier $\frac{6}{5}$ is sufficient
 to verify (\ref{fsf}).

    Define next the $(63\times 63)$ matrix
 \begin{align*}
   {\bf A}(t)& :=\frac{D {\bf f}(u^\ast(t))}{D \bf u^\ast} \,, \quad
      A_{ms}(t) =\frac{\partial f_m(u^\ast(t))}{\partial u^\ast},
     \quad    m,s=1,\ldots,63 \ ,
 \end{align*}
 as well as the fundamental solution
 $\boldsymbol\Phi=\boldsymbol\Phi(t)$ for the linear
 system of $63^2$ ODE's
 \begin{align*}
   \boldsymbol {\dot \Phi} ={\bf A}(t)\boldsymbol \Phi , \qquad
   {\boldsymbol\Phi}(0)= {\bf I}\,,
 \end{align*}
 ${\bf I}$ is the identity matrix.
 Remark that ${\bf A}(t)$ has been already known and
 $\boldsymbol\Phi$ can be computed approximately as follows.
 We start-up the program $64$ times independently of one another.
 At the $m$th start-up we change only the $m$th initial
 condition to be
 \ben
   \big( u_1^*(0),\ldots,u_{m-1}^*(0),u_{m}^*(0)+10^{-5},
         u_{m+1}^*(0),\ldots,u_{63}^*(0) \big)
 \een
 and thus we find the $m$th column of $\boldsymbol\Phi$.
 The change in the mass of the Sun gives the final $\boldsymbol\Phi$.

   We perform these additional numerical integrations and conclude
 that $\boldsymbol\Phi$ remains close to $\boldsymbol\Phi(0)$:
 \be
  \label{Phi}
   \left\lVert \boldsymbol \Phi(t)-{\bf I} \right\rVert < \frac{1}{5}
 \ee
 for all $t < 2\pi 10^5$ and the matrix norm associated with
 the vector norm (\ref{vnorm}).

 Now we compute
 \be
  \label{derdev}
   \dot{\boldsymbol\delta}(t)
   & = {\bf f}({\bf v}(t))-{\bf f}^\ast({\bf u}^\ast(t))   \\[4pt]
   & = {\bf f}^\ast({\bf u}^\ast+{\boldsymbol\delta})
      -{\bf f}^\ast({\bf u}^\ast)+{\bf f}({\bf v})
      -{\bf f}^\ast({\bf v})                                \\[4pt]
   & = {\bf A}(t){\boldsymbol\delta}(t)+{\bf F}(t)+{\bf G}(t)\,,
 \ee
 where the functions
 \begin{align*}
     {\bf F}(t) & := \dot{\bf u}^*(t)-{\bf A}(t){\bf u}^*(t)   \,,
  \\[4pt]
     {\bf G}(t) & := {\bf f}({\bf v}(t))-{\bf f}^*({\bf v}(t)) \,,
 \end{align*}
 will be estimated in the next subsection.

 The weakly non-linear system (\ref{derdev}) satisfies \citep{Cesari}
 the non-linear Volterra integral equation
 \be
   \label{Vol}
       {\boldsymbol\delta}(t)
       &= \boldsymbol\Phi(t)
         \biggl[  {\boldsymbol\delta}(0)
       + \int_{0}^{t}{\boldsymbol\Phi}^{-1}(\alpha)
       [{\bf F}(\alpha)+{\bf G}(\alpha)]\,d\alpha \biggr]
 \ee
and our aim is to prove that $\left\lVert {\boldsymbol\delta}(t)
\right\rVert$ is sufficiently small for all $0 \le t \le 2 \pi \cdot
10^5$.

 \subsection{End of the proof of the main Theorem}
 \label{EndTh}

 According to (\ref{fsf}), $||{\bf G}||<5\cdot\!10^{-10}$
 provided $||{\bf v}-{\bf u}^*||<10^{-3}$
 for all $t$.

 More complicated is to estimate the vector function ${\bf F}$.
 To do this we analyze the second order reminder in
 the Taylor's expansion of
 ${\bf f}^\ast({\bf u}^\ast+{\boldsymbol\delta})$,
 by definition equal to ${\bf F}$:
 \begin{align}
  \label{Fm}
     F_m(t)
   &:= \frac{1}{2}
        \sum\limits_{b=1}^{54}
        \sum\limits_{c=1}^{54}
        \frac{\partial^2 f^*_m(\boldsymbol\xi (t))}
             {\partial u^*_b \,\partial u^*_c }
        \delta_b \delta_c \ ,
 \end{align}
 where $F_m$ are the components of the vector ${\bf F}$,
 $1\leq m \leq 54$ and
 $||{\boldsymbol\xi}-{\bf u}^*|| < ||{\bf v}-{\bf u}^*||$
 for all $t$.

   Next we estimate the second partial derivatives
 in (\ref{Fm}) by
 \begin{align*}
     F_{m;b,c}
     &:= \sup\limits_t
          \Biggl| \frac{\partial^2 f^*_m(\boldsymbol\xi (t))}
                       {\partial u^*_b \,\partial u^*_c }
          \Biggr| \ .
 \end{align*}
 Every index $m,b$ or $c$ corresponds to an orbital element
 or a relative mass.
 We shall distinguish the following three types of $F_{m;b,c}$:

  (i) vanishing $F_{m;b,c}$ -- if $m,b$ and $c$ correspond
  to three different planet, or if some of them corresponds
  to some relative mass,

  (ii) negligible small $F_{m;b,c}$ -- if  $m,b$ or $c$
  corresponds to an element of any inner planet or Pluto
  (their masses are too small); or if $m,b$ or $c$
  corresponds to some oblate orbital element,

  (iii)  $F_{m;b,c}$ is neither of type (i) nor of type (ii).
  For any fixed $m$ there exist $16$ such orbital elements:
  $a$, $\epsilon$, $h$ and $k$ of Jupiter, Saturn,
  Uranus and Neptune, and always
  \begin{align*}
     F_{m;b,c} < \frac{1}{3000} \ .
  \end{align*}

     All these examinations together enable us to
  conclude that
  \ben
     ||{\bf F}|| < \frac{||\boldsymbol\delta||^2}
                        {1000}  \ .
  \een
  Compared with (\ref{Phi}), (\ref{Vol}) and the
  already made estimation of ${\bf G}$ this proves
  that
  \begin{align*}
     ||\boldsymbol\delta (t)|| \,
     &< \
     \frac{3}{2}
     \int\limits_0^t
        \biggl[ \frac{||\boldsymbol\delta(t)||^2}{1000}
                + 5\!\cdot\!10^{-10}
        \biggr] dt
  \\
     &\, = \ \frac{\sqrt{2}}{2000}
         \tan \biggl( \frac{3\sqrt{2}\,t}{4\cdot\!10^6}
                      +\arctan\frac{\sqrt{2}}{10}     \,
              \biggr).
  \end{align*}

    Hence $||\boldsymbol\delta (t)||$ is sufficiently small
 for $0 \leq t \leq 2\pi 10^5$, satisfies the requirement
 (\ref{sigma}) and the theorem has been finally proved.

 \section{Conclusions}
 \label{Conclusions}

 Throughout the proof of the Theorem, we have not used
 the fact that for long periods of time the perturbations
 of the orbital elements are canceling almost entirely.

   With simple additional reasonings and evaluations,
 while preserving the numerical integration algorithm,
 the Theorem could be proven for one million years.

   Presumably under more detailed additional evaluations
 and availability of powerful computer,
 the stability of the solar system could be proved
 for the next $5\! \cdot\! 10^9$ years.

   We will note also that the evaluations which are done
 about the deviation between the numerical integration
 and the real dynamics are a little rough.
 Hence it is possible to include in them additional real
 existing details such as:
 \begin{itemize}
 \item relativistic effects
 \item the fact that the planets are not point masses
 \item passing of external sufficiently small celestial
       bodies across the solar system.
 \end{itemize}
   However accounting the influence of the known small
 bodies in the solar system demands huge extra labor.

  The numerical integration of the solar system dynamics
 indicates that the fourth order secular terms should be taken
 into consideration in the purely analytical
 approaches.

 \section*{Acknowledgements}

 The authors acknowledge support from Bulgarian Fund
 ``Scientific Research'' through grant DN 02-5
 and from Sofia University ``St. Kliment Ohridski''
 through grant 80-10-5/18.03.2021.

 \end{document}